# M-SGWR: Multiscale Similarity and Geographically Weighted Regression

M. Naser Lessani[1], Zhenlong Li[1*], Manzhu Yu[2], Helen Greatrex[3], and Chan Shen[4]

[1]Geoinformation and Big Data Research Laboratory, Department of Geography, The Pennsylvania State University, University Park, PA, USA

[2]Department of Geography, The Pennsylvania State University, University Park, PA, USA

[3]Department of Geography, Department of Statistics and the Institute for Computational and Data Sciences, The Pennsylvania State University, University Park, PA, USA

[4]Departments of Surgery and Public Health Sciences, Penn State College of Medicine, 400 University Drive, Hershey, PA 17033, USA

[*]Corresponding author: zhenlong@psu.edu

**Abstract:** The first law of geography is a cornerstone of spatial analysis, emphasizing that nearby and related locations tend to be more similar; however, defining what "near" and "related" mean remains challenging, as different phenomena operate across distinct spatial dimensions (e.g., physical, social, or network-based space). Traditional local regression models, such as Geographically Weighted Regression (GWR) and Multiscale GWR (MGWR), quantify spatial relationships solely through geographic proximity. In an era of globalization and digital connectivity, however, geographic proximity alone may be insufficient to capture how locations are interconnected. To address this limitation, we propose a new multiscale local regression framework, termed M-SGWR, which characterizes spatial interaction across two dimensions: geographic space and attribute (variable) space. For each predictor, geographic and attribute-based weight matrices are constructed separately and then combined using an optimized parameter, $\alpha$, which governs their relative contribution to local model fitting. Analogous to variable-specific bandwidths in MGWR, the optimal $\alpha$ varies by predictor, allowing the model to flexibly account for geographic, mixed, or non-spatial (remote similarity) effects. Results from two simulation experiments and one empirical application demonstrate that M-SGWR consistently outperforms GWR, SGWR, and MGWR across all goodness-of-fit metrics.

1. Introduction

Few processes unfold evenly across space. Instead, spatial relationships reflect the influence of place-specific conditions and the uneven geographies in which they occur. Despite this, traditional global models such as Ordinary Least Squares (OLS) assume spatial stationarity, applying constant regression parameters across entire study regions (Sykes, 1993; Wooditch et al., 2021). This limitation has driven the advancement of local regression models, which relax the assumption of global stationarity and allow regression coefficients to vary across space (A. S. Fotheringham et al., 2003; Brunsdon et al., 1996, 2010). As a result, they are increasingly applied in geographic information science and related fields, including public health, criminology, housing price modeling, environmental science, ecology, and geoscience (Cao et al., 2018; Luo et al., 2017; Nezami & Khoramshahi, 2016; Wu, 2020). A variety of techniques have been proposed to operationalize local regression estimates, including Geographically Weighted Regression (GWR), spatial filtering, Bayesian spatially varying coefficient (SVC) regression, moving window regression, eigenvector spatial filtering (ESF)–based SVC models (Yang, 2014). In recent years, the integration of neural networks and Spatial Process Transfer Learning into regression model has also emerged as a promising approach for capturing complex spatial structures (Du et al., 2020; Hagenauer & Helbich, 2021; Murakami et al., 2025). Each of these local regression models seeks to address spatial dependency and spatial heterogeneity in its own unique way.

One of the most widely used approaches is GWR, which models spatial heterogeneity by fitting a separate local regression at each observation point (Brunsdon et al., 2010). It assigns spatial weights to neighboring data using a kernel function defined by a single bandwidth parameter, so that nearer observations receive higher weights than more distant ones, reflecting the First Law of Geography (A. S. Fotheringham et al., 2003; Tobler, 1970). Although the original formulation employs Euclidean distance, more recent implementations incorporate non-Euclidean metrics to account for anisotropy and spatial irregularities, including travel times, road-network distances, Minkowski metrics, and hydrological networks (Curriero, 2006; Lu et al., 2014; Lu et al., 2015; Mainali et al., 2022). Despite these advancements, a key limitation of standard GWR is that it applies the same bandwidth to all predictors, even though different variables may operate across different spatial scales (Wolf et al., 2017; Yang, 2014).

To address this, researchers developed Multiscale GWR (MGWR), which was first introduced in Mixed GWR, where some relationships are assumed to be stationary (globally fixed) while others are treated as non-stationary (spatially varying) (Brunsdon et al., 1999). However, Mixed GWR only partially overcomes the single-scale limitation, as some spatially varying relationships are still assumed to operate at the same scale. Yang et al. later expanded this idea by proposing a unique spatial scale (bandwidth) for each variable in the model, allowing each relationship to be examined at its own scale, an approach termed Flexible Bandwidth GWR (FBGWR) (Yang, 2014). Building on this, Multiscale GWR (MGWR) was later developed using Euclidean distance to construct the spatial weight matrix (Fotheringham et al., 2017). Lu et al. (2015, 2017) proposed a similar multiscale concept but adding non-Euclidean distance metrics in the model in addition to Euclidean distance, resulting in the GWR with Parameter-Specific Distance Metrics (PSDM GWR) model (Lu et al., 2016; Lu et al., 2015). The aforementioned models draw on the concept of Generalized Additive Models (GAM). Thus, they are computationally demanding due to the need for multi-bandwidth optimization and model fitting at each location. Faster versions of these models have been developed to overcome the computational time of some of these models (Li et al., 2019; Li & Fotheringham, 2020; Lu et al., 2018; Wu et al., 2021; Zhou et al., 2023).

The models discussed thus far rely exclusively on geographic space (physical distance) to quantify spatial dependency between pairs of spatial units (Lessani & Li, 2024). While this simplifies estimation, it fails to account for the dynamic and multidimensional nature of space, which is shaped not only by physical separation but also by perception, context, and interactions that change over time (Brockmann & Helbing, 2013; He et al., 2023). Empirical examples underscore the limits of geographic proximity as a proxy for closeness. For instance, the global diffusion of infectious diseases such as influenza is better predicted by effective distances on airline networks than by Euclidean distance (Brockmann & Helbing, 2013; Lessani et al., 2023; Unwin, 1996). Likewise, online social networks show that digital connectivity, shared language, and air-travel routes often predict social ties more strongly than physical location (Takhteyev et al., 2012). Given these complexities, models that represent spatial dependence solely through fixed geographic distance may misrepresent the structure of spatial interactions, leading to biased estimates and inference (Foresman & Luscombe, 2017; Tobler, 2004; Tobler, 1970).

To address this limitation, recent studies have introduced an additional dimension for measuring spatial dependency alongside geographic space. This line of research is based on the idea that if two locations share similar geographic configurations, their level of interaction may be greater, and vice versa (Zhu & Turner, 2022). Lessani and Li proposed a new local regression model, termed Similarity and Geographically Weighted Regression (SGWR), which measures proximity between locations across two dimensions—geographic distance and attribute similarity—and combines these measures through a tuning parameter ($\alpha$) (Lessani & Li, 2024). Similar concepts have been applied in spatial prediction modeling, such as incorporating contextual similarity to estimate non-stationary processes, predicting soil maps based on geographic configuration similarities, groundwater level change, and integrating geographic similarity into multivariate spatial structure indicators (Wu et al., 2025; Zhao et al., 2023; Zhao et al., 2025; Zhu et al., 2018). We argue that data (attribute) space and geographical (physical) space are complementary dimensions of spatial proximity, rather than independent or competing measures.

While SGWR model has enhanced the explanatory power and predictive accuracy of local spatial models, it still operates under a single spatial scale (single bandwidth). Additionally, because a single tuning parameter ($\alpha$) is applied across all predictors, the model cannot distinguish whether individual predictors are governed primarily by geographic proximity, attribute similarity (e.g., social networks), or a combination of both. As noted earlier, spatial phenomena often occur at multiple scales, and assuming a uniform scale can lead to misrepresentation of spatial patterns. To address this limitation, the present study extends the SGWR framework from a single-scale to a multiscale approach. In Multiscale Similarity-Geographically Weighted Regression (M-SGWR), we extend SGWR by assigning a distinct ($\alpha$) value and distinct spatial scale (bandwidth) to each predictor. Analogous to the role of variable-specific bandwidths in MGWR, our model allows each coefficient to learn its own balance between spatial and attribute-based weighting. This acknowledges that some processes may be more spatially structured, while others are more influenced by relational similarity. By aligning the model's spatial structure with the generative logic of each predictor, M-SGWR aims to improve model fit, inference, and interpretability. Given the widespread use of GWR and MGWR in domains such as housing, health, ecology, and environmental risk, M-SGWR offers an actionable extension with immediate relevance to applied spatial modelling.

## 2. Methodology

### 2.1. Similarity and geographically weighted regression (SGWR)

The SGWR model constructs a local regression for each observation by selecting samples from neighboring locations by quantifying the proximity of these samples to the regression point through both a spatial weight matrix and an attribute weight matrix. The core principle of the SGWR model is that regions with similar attributes tend to exhibit stronger interactions, with 'closeness' and 'relatedness' measured across two dimensions: geographical space and attribute (variable) space. The mathematical expression of the SGWR model can be written as follows:

$$y_i = \beta_0(u_i, v_i) + \sum_{j=1}^{m} \beta_j(u_i, v_i)X_{ij} + \varepsilon_i, \quad i = 1,2,3,\ldots,n \quad (1)$$

$$\beta_j(u_i, v_i) = [X^T W_{GS}(u_i, v_i) X]^{-1} X^T W_{GS}(u_i, v_i) y \quad (2)$$

$$W_{GS}(u_i, v_i) = \alpha * W_G(u_i, v_i) + \gamma * W_S(u_i, v_i) \quad (3)$$

where $y_i$ denotes the value of the response variable at location $(i)$; $\beta_0(u_i, v_i)$ is the estimated intercept at location $(i)$ with coordinates $(u_i, v_i)$; $\beta_j(u_i, v_i)$ is the estimated coefficient for the $(j)$ predictor at location $(i)$; $X_{ij}$ is the value of the $j$-th independent variable at location $(i)$; and $\varepsilon_i$ is the corresponding error term. Here, $m$ and $n$ represent the number of independent variables (including the intercept) and the total number of observations, respectively. $W_{GS}$ denotes the integrated weight matrix, formed by combining the spatial weight matrix ($W_G$) and the attribute weight matrix ($W_S$). The parameter $\gamma$ is $(1 - \alpha)$, and alpha ($\alpha$) ranges from zero to one, which controls the relative influence of spatial and attribute weights: values of ($\alpha$) closer to one assign greater importance to spatial proximity, while values closer to zero place more emphasis on attribute similarity.

### 2.2. Multiscale SGWR model

In the multiscale SGWR model, hereafter referred to as M-SGWR, the multiscale concept is extended such that each independent variable is assigned its own optimal alpha ($\alpha$) value along with a distinct bandwidth (Fotheringham et al., 2017; Yang, 2014; Yu et al., 2019). Bandwidth is a parameter that controls how much of the surrounding data is used to fit the local regression at each location. A smaller bandwidth for a given variable indicates greater local variation, meaning the variable exhibits more spatial heterogeneity. Conversely, a larger bandwidth suggests smoother spatial variation, where the influence of neighboring observations decays more gradually with distance, resulting in smoother coefficient surfaces. In essence, the bandwidth controls the spatial pattern of the relationship, and its magnitude reflects the scale at which the underlying spatial process operates. We argue that analogous to the use of multiple bandwidths, incorporating multiple ($\alpha$) values allow the model to capture spatial interactions between location pairs more accurately and to better represent real-world scenarios in attribute (variable) space. This enables assessment of how each individual variable in a local regression model is influenced by both geographical proximity and attribute similarity (proximity). Instead of applying a single ($\alpha$) value to all predictors as in SGWR model, a distinct ($\alpha$) is assigned to each independent variable, reflecting the relative contribution of physical distance and attribute similarity for each variable. For example, in predicting the amount of fruit harvested from a tree using water availability and sunlight exposure as predictors, the M-SGWR model may assign ($\alpha = 0.8$) for water and ($\alpha = 0.6$) for sunlight. This would indicate that the spatial effect of water is more strongly governed by geographical distance, whereas the influence of sunlight is more closely associated with attribute similarity of the selected observations in the selected neighborhoods. As illustration, Figures 1 and 2 are presented to demonstrate how proximity might differ across two dimensions and how the outcome variable can be measured.

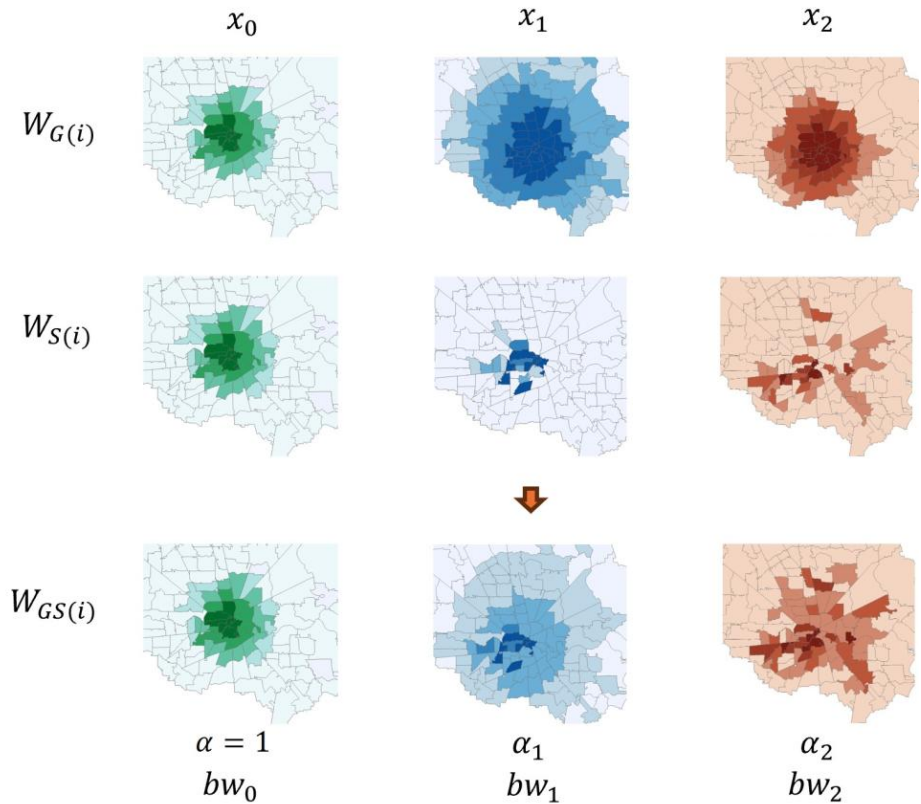

**Figure 1.** M-SGWR weight generation per observation with variable-specific α and bandwidths.

Figure 1 is illustrated for a better visualization of M-SGWR model, the attribute weight for the intercept ($x_0$) equals the geographic weight because $\alpha$ is set to 1, given that all intercept values are identical. For other variables, however, the attribute weights vary because of their distinct $\alpha$ values. As expected, spatial weights follow a smooth kernel decay, whereas attribute-based weights do not follow a kernel function. Instead, they capture variations that spatial weights overlook because they are not constrained by geographic distance decay. Figure 2 illustrates how the response variable values are generated based on different optimal $\alpha$ and bandwidths in the model. As also shown in Figure 2, the estimated coefficients in M-SGWR may appear less smooth than in the MGWR model, since M-SGWR captures more local variation. This will be further discussed in Section 3.5. The circles in Figure 2 indicate the size of bandwidths.

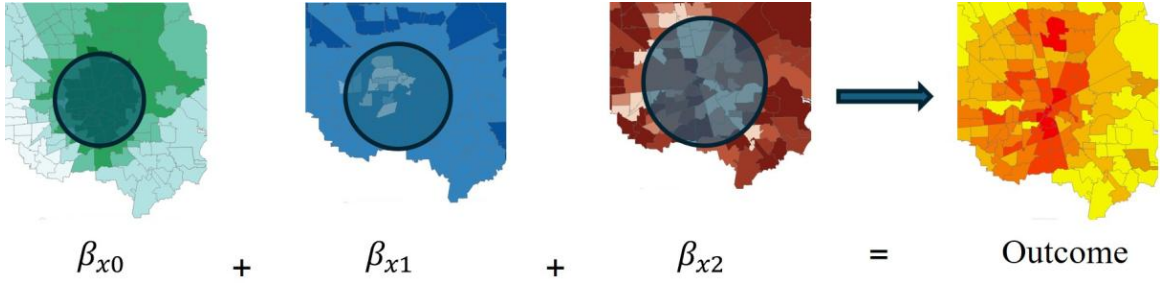

$\beta_{x0}$ + $\beta_{x1}$ + $\beta_{x2}$ = Outcome

**Figure 2.** Prediction of outcome variable based on M-SGWR model.

2.3. Mathematical expressions of M-SGWR model

M-SGWR can be expressed mathematically as follows: where $(bwj)$ in $\beta_{bwj}^{(j)}$ denotes the bandwidth applied in calibrating the $j$-th conditional relationship, $\alpha^{(j)}$ is the unique mixing ($\alpha$) value for the $j$-th variable, $W_{GS}^{(j)}$ is the mixed weight matrix for variable $j$ obtained by combining the geographically weighted matrix $W_G^{(j)}$ and attribute similarity weight matrix $W_S^{(j)}$.

$$y_i = \sum_{j=0}^{k} \beta_{bwj}^{(j)}(u_i, v_i) X_{ij} + \varepsilon_i, \quad i = 1,2,3,\ldots,n \tag{4}$$

$$\beta_{bwj}^{(j)}(u_i, v_i) = \left[X^T W_{GS}^{(j)}(u_i, v_i) X\right]^{-1} X^T W_{GS}^{(j)}(u_i, v_i) y \tag{5}$$

$$W_{GS}^{(j)}(u_i, v_i) = \alpha^j W_G^{(j)}(u_i, v_i) + (1 - \alpha^j) W_S^{(j)}(u_i, v_i) \tag{6}$$

Currently, in M-SGWR model, adaptive bi-square kernel (Equation 7) functions is available, where $d_{i\ell}$ is the distance between location $(i)$ and $(\ell)$ and $bw_j$ denotes the bandwidth for variable $(j)$. The procedure of generating the attribute weight matrix will be discussed in Section 2.4.

$$W_G^{(j)}(i,\ell) = \begin{cases} \left[1 - \left(\frac{d_{i\ell}}{bw_j}\right)^2\right]^2, & d_{i\ell} < bw_j \\ 0, & Otherwise \end{cases} \tag{7}$$

Equation (8) defines the local weighted least squares smoothing operator $A_j$ for covariate $j$ at regression location $i$. Here, $W_{GS,i}^{(j)}$ denotes the mixed geographic–similarity weight matrix for covariate $j$ when location $i$ is the regression point, and $x_{ij}$ is the value of covariate $j$ at location $i$. Importantly, $A_j$ is not the model hat matrix, instead, it is used within the backfitting procedure to iteratively construct the covariate-specific projection matrix $R_j$, following the MGWR inference framework of Yu et al. (2019).

$$[A_j]_i = x_{ij}(X_j^T W_{GS,i}^{(j)} X_j)^{-1} X_j^T W_{GS,i}^{(j)} \tag{8}$$

The covariate specific effective number of parameters ($ENP_j$) can be written as in Equation 9, where $R_j$ is the partial projection matrix for variable ($j$) after backfitting (backfitting algorithm will be discussed in Section 2.5); Equation 10 shows the overall ENP for the model, where $S = \sum_{j=1}^{k} R_j$, and $j = 0$ if we include the intercept as well.

$$ENP_j = trace(R_j) \tag{9}$$

$$ENP_{model} = trace(S) \tag{10}$$

For obtaining standard errors, the model first estimates the residual variance (Equation 11), where $y_i$ denotes the observed value of the response variable and $\hat{y}_i$ denotes the fitted value produced by the fully converged M-SGWR model at location $i$.

$$\hat{\sigma}^2 = \frac{\sum_{i=1}^{n}(y_i - \hat{y}_i)^2}{n - trace(S)} \tag{11}$$

Then, Equation (12) gives the variance of the local coefficient estimate for covariate $j$ at location $i$. The variance is obtained by propagating the residual variance $\hat{\sigma}^2$ through the covariate-specific projection matrix $R_j$, scaled by the local values of covariate $j$.

$$\text{var}\left(\beta^{(j)}(u_i, v_i)\right) = \text{diag}\left(([\text{diag}(X_j)]^{-1} R_j)([\text{diag}(X_j)]^{-1} R_j)^T \hat{\sigma}^2\right) \tag{12}$$

Equation (13) computes the location-specific standard error of the estimated coefficient for covariate $j$ as the square root of its variance. Equation (14) then forms a pseudo-t statistic by standardizing the local coefficient estimate with its standard error, providing an approximate measure of local coefficient significance under the multiscale backfitting framework.

$$SE\left(\beta_{bwj}^{(j)}(u_i, v_i)\right) = \sqrt{\text{var}\left(\beta^{(j)}(u_i, v_i)\right)} \tag{13}$$

$$t_{(i,j)} = \frac{\beta_{bwj}^{(j)}(u_i, v_i)}{SE\left(\beta_{bwj}^{(j)}(u_i, v_i)\right)} \tag{14}$$

2.4. The attribute similarity-based weight matrix

In the SGWR model, attribute similarity between pairs of locations is calculated using all predictors to obtain one similarity score between two locations. For each variable, the pairwise distance between the two locations is computed separately, and these distances are then averaged to obtain a single similarity score based on exponential decay function. In M-SGWR, the similarity between a predictor's values at two locations is expressed on a scale from 0 to 1 similar to SGWR model. However, the attribute weighting procedure partially follows Zhao et al., method (Zhao et al., 2023). Here, for each variable ($j$) and target location ($i$), let $\mathcal{N}_i = \{\ell : W_G^{(j)}(i, \ell) > 0\}$ be the set of geographically selected neighbors (other locations get zero attribute weight).

$$SD_{\mathcal{N}_i}^{(j)} = std\{x_1^{(j)}, \ldots, x_n^{(j)}\}, \quad \widetilde{SD}_{\mathcal{N}_i}^{(j)} = \begin{cases} SD_{\mathcal{N}_i}^{(j)}; & SD_{\mathcal{N}_i}^{(j)} > 0 \\ \rho, & SD_{\mathcal{N}_i}^{(j)} = 0; \, (\rho \approx 10^{-5}) \end{cases} \quad (15)$$

$$W_S^{(j)}(i, \ell) = \begin{cases} 0.5^{\left(\frac{x_\ell^{(j)} - x_i^{(j)}}{\widetilde{SD}^{(j)}}\right)^2}, & \ell \in \mathcal{N}_i \\ 0, & \ell \notin \mathcal{N}_i \end{cases} \quad (16)$$

$$W_{S,i}^{(j)} = diag\left(W_S^{(j)}(i, \ell), \ldots, W_S^{(j)}(i, n)\right) \quad (17)$$

In Equation 16, $x_i^{(j)}$ and $x_\ell^{(j)}$ are the predictor value of ($j$) at location ($i$) and ($\ell$), respectively; $SD^{(j)}$ is the standard deviation of the predictor ($j$) in the neighborhood; $\widetilde{SD}^{(j)}$ denotes the adjusted $SD^{(j)}$, if the standard deviation happens to be zero, it is replaced with a very small constat ($\rho \approx 10^{-5}$) to avoid division by zero in the weight formula, otherwise it will receive the same $SD^{(j)}$. $W_S^{(j)}(i, \ell)$ is the attribute similarity weight matrix for variable ($j$) between location ($i$) and ($\ell$). $W_{S,i}^{(j)}$ is a diagonal weight matrix and it denotes the weight matrix used when location ($i$) is the regression point (Equation 17). Here, the similarity is determined using a Gaussian-shaped curve, whose width is scaled by the standard deviation ($SD_{\mathcal{N}_i}^{(j)}$). This scaling ensures that when the difference between two values equals the $SD$, the resulting similarity is 0.5.

## 2.5. Bandwidth and alpha optimization

Bandwidth plays a critical role in local regression models, as it governs the degree of spatial localization by determining the subset of observations used in calibrating the model at each location. In the GWR model, the kernel function specifies how spatial weights decay by increasing geographical distance from the regression point, thereby controlling the influence of distant observations on local parameter estimates. The choice of kernel such as Gaussian, adaptive bisquare, exponential, or tricube affects the smoothness, range, and sensitivity of the weighting scheme, with each kernel exhibiting distinct strengths and limitations. To identify the optimal bandwidth, several evaluation metrics can be utilized, including the AICc and cross-validation (CV) (Fotheringham et al., 2017; Yang, 2014). In this study, AICc and CV measures are available for bandwidth and alpha optimization, and they are defined as Equations 18 and 19, respectively:

$$AIC_C = n \log_e(\hat{\sigma}^2) + n \log_e(2\pi) + n\left(\frac{n + tr(S)}{n - 2 - tr(S)}\right) \quad (18)$$

$$CV\ score = \frac{1}{n}\sum_{i=1}^{n}\left(\frac{e_i}{1 - s_{ii}}\right)^2 \quad (19)$$

In Equation 18, ($\hat{\sigma}$) denotes the estimated standard deviation of the error term, $tr(S)$ is the trace of the hat matrix ($S$), and ($n$) is the number of observations. For Equation 19, $e_i$ is the difference between the observed and predicted value of location $i$ and $s_{ii}$ is the *ith* diagonal of the hat matrix. Conceptually, this approach is equivalent to leave-one-out cross-validation (LOOCV), in which the model is trained on all observations except one that is held out for prediction. The process is repeated for every observation in the dataset so that each point takes a turn as the "left-out" case. In our model, akin to GWR and MGWR, this procedure is not performed by literally refitting the model *n* times, doing so would be computationally prohibitive, as each trial bandwidth and α value would require *n* separate model fits. Instead, the computation is achieved through a mathematical shortcut. Specifically, the leave-one-out residual for each observation is approximated as ($e_i/(1 - s_{ii})$), where $e_i$ is the residual from the full model and $s_{ii}$ is the leverage value representing how much observation $i$ influences its own prediction (Equation 19). By applying this correction, the algorithm efficiently estimates the error that would have been obtained if the observation had been excluded. Finally, all squared leave-one-out residuals are summed and averaged to produce the overall cross-validation score, which measures how well the model generalizes to unseen data across all locations. The optimal bandwidth is the one that minimizes the AICc or CV value. Bandwidth selection is therefore an optimization procedure aimed at finding the bandwidth that minimizes the chosen diagnostic measure within a predefined search range.

The alpha ($\alpha$) parameter is another critical component of the M-SGWR model, as it controls the relative contribution of the spatial and attribute weight matrices in forming the combined weight matrix ($W_{GS}^{(j)}$), thereby influencing the model's goodness-of-fit. In essence, ($\alpha$) determines the balance between spatial proximity and attribute similarity in the weighting scheme for each variable. During the backfitting process for bandwidth optimization, ($\alpha$) is optimized within the range of [0,1] for each candidate bandwidth. In this study, two search strategies are implemented to determine the optimal ($\alpha$): (1) a divide-and-conquer approach with adaptive step sizes, and (2) a greedy hill-climbing approach with fixed step sizes initiated from a set of predefined seed values. The first method is more computationally intensive than the greedy method, but it can yield a more accurate alpha value because its step size is determined adaptively during the search, whereas in the greedy method, the step sizes are predefined within the model. To identify the optimal ($\alpha$) for each candidate bandwidth, we can select either the AICc or CV metric, selecting the ($\alpha$) value that yields the minimum AICc value or lower CV score. Each optimal ($\alpha$) corresponding to a tried bandwidth is recorded during the bandwidth optimization process and subsequently used in the model inference stage.

### 2.6. Backfitting algorithm

Multi-bandwidth optimization in MGWR constitutes a combinatorial search problem, as it involves exploring all possible combinations of variable-specific bandwidths; $(b_1, b_2, \ldots, b_k) \in \mathcal{B}_1 \times \mathcal{B}_2 \times \ldots \times \mathcal{B}_k$, where $b_j$ is the bandwidth for the $j$-th predictor variable in multiscale GWR and $\mathcal{B}_j$ is the set of candidate bandwidth values considered for $j$-th variable during optimization. Consequently, metaheuristic algorithms such as Genetic Algorithms or Simulated Annealing could be employed to navigate this search space; however, these methods are typically computationally demanding due to the large number of model evaluations required. Even if the model uses continuous search methods (Golden Section Search and gradient-based methods), the underlying search space still consists of all possible combinations of bandwidths across variables. The backfitting algorithm is essentially breaking down this combinatorial search into iterative one-dimensional searches, which makes it tractable. This study also utilizes a backfitting algorithm instead of metaheuristic optimizers because of their computational efficiency and stability. Backfitting exploits the fact that the multiscale objective surface is often smooth and unimodal in each variable's bandwidth dimension, so a one-dimensional optimizer like Golden Section Search converges faster.

The first step in the backfitting algorithm is initialization (Figure 3). First, set initial bandwidth $b_j^{(0)}$ value for each variable $j$; second, initialize coefficient vectors $\beta_j^{(0)}$, which we used SGWR model; third, set convergence tolerance ($\varphi$) and the number of iterations. The second stage is backfitting iteration, for each iteration and then for each variable ($j = 1, 2, ..., k$: 1) partial residual is computed, which remove the contribution of $j-th$ predictor from the model using Equation 20 and here ($\odot$) denotes element-wise multiplication; by removing the contribution of $j-th$ variable from the model we can isolate the effect of a single variable when estimating its local bandwidth while holding all other variables fixed. Then for each tried bandwidth ($b\epsilon\ \mathcal{B}_j$), the model optimizes an ($\alpha$) in the range of [0, 1] using either divide and conquer or greedy hill-climb method either utilizes AICc or CV measures (Equation 21 and 22); then store the pair ($bw_j, \alpha^*_{bw(j)}$) and the corresponding AICc value here. Next, select $b_j^{(t+1)}$ in Equation 23 and with the optimal ($\alpha$) value from Equation 21. Then calculate local weight for each location based on variable $j$ utilizing Equation 23 and subsequently estimate the local coefficients using Equation 24; for location $i$, where $W_{GS,i}^{(j)}$ is the diagonal weight matrix for location $i$ based on predictor $j$. Note that the asterisk in Equation 21 denotes the optimal $\alpha$ value for variable $j$ based on the bandwidth selected for that predictor.

$$r^{(j)} = y - \sum_{l \neq j}^{k} \beta_l^{(t)} \odot x_l \tag{20}$$

$$\alpha^*_{bw(j)} = \arg\min_{\alpha\epsilon[0,1]} AICc(bw_j, \alpha) \tag{21}$$

$$bw_j^{(t+1)} = \arg\min_{b\epsilon\ \mathcal{B}_j} AICc(bw_j, \alpha^*_{bw(j)}) \tag{22}$$

$$W_{SG}^{(j)} = \alpha^*_{bw(j)} W_G^{(j)}(bw_j^{t+1}) + (1 - \alpha^*_{bw(j)}) W_S^{(j)} \tag{23}$$

$$\beta_j^{(t+1)}(u_i, v_i) = (X_j^T W_{GS,i}^{(j)} X_j)^{-1} X_j^T W_{GS,i}^{(j)} r^{(j)} \tag{24}$$

After updating all variables, the model checks its convergence, which is the choice of termination criterion. Following the MGWR framework, this study allows the model to terminate based on either the change in estimated coefficients between the current and previous iterations (Equation 25) or the residual sum of squares (26) (Fotheringham et al., 2017). In Equation 25, $X\mathcal{B}_{old(ij)}$ fitted contribution of predictor ($j$) for observation ($i$) from the previous iteration, and $X\mathcal{B}_{new(ij)}$ is fitted contribution of predictor ($j$) for observation ($i$) in the current iteration. In Equation 26, $RSS_{new}$ and $RSS_{old}$ are the residual sum of squares from the current and previous iterations, respectively.

$$SOC = \sqrt{\frac{\sum_{j=1}^{k} \frac{\sum_{i=1}^{n}(X\mathcal{B}_{new(ij)} - X\mathcal{B}_{old(ij)})^2}{n}}{\sum_{i=1}^{n}(\sum_{j=1}^{k} X\mathcal{B}_{new(ij)})^2}} \quad (25)$$

$$SOC_{RSS} = \frac{|RSS_{new} - RSS_{old}|}{RSS_{new}} \quad (26)$$

In the inference stage, the optimal bandwidth and alpha values are no longer being searched or adjusted but they are fixed. For each variable, the model uses its own optimal bandwidth to calculate a spatial weight matrix and its own optimal ($\alpha$) value to blend this with the attribute similarity weight matrix. This produces a combined weight matrix that reflects the best balance between spatial closeness and similarity in predictor values. The model then goes through each location in the study area. At each location, it uses the combined weight matrices concept for all variables to run a local regression that estimates the coefficients at that specific point. Because each variable has its own bandwidth and ($\alpha$) value, the model operates at multiple scales in both spatial and attribute dimensions. This allows the model to determine which variables are more influenced by spatial proximity and which are more strongly driven by attribute similarity (attribute proximity).

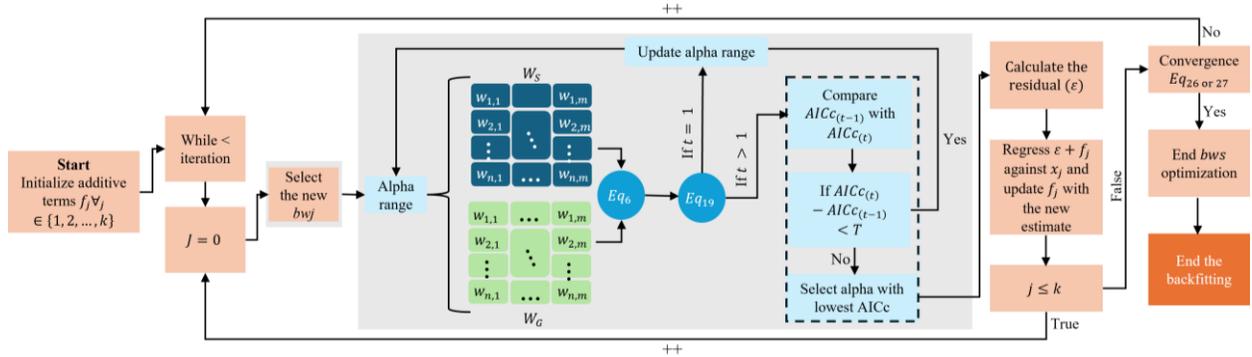

**Figure 3.** M-SGWR framework.

It is important to clarify that the attribute-based weight matrix in M-SGWR does not have an associated spatial bandwidth. As described in Section 2.4, attribute similarity is computed between the regression location and its geographically defined neighbors and is normalized by the standard deviation within that neighborhood. Consequently, the geographic and attribute weight matrices do not share a common bandwidth: geographic weights are governed by a distance-based bandwidth, while attribute weights are only normalized (Equation 16). During the backfitting, geographic bandwidths are optimized following the MGWR procedure, and for each candidate bandwidth an optimal mixing parameter, $\alpha$, is jointly estimated to control the relative contribution of geographic and attribute-based weights. This design allows geographic proximity and attribute similarity to operate jointly yet independently.

## 3. Simulation design

The M-SGWR model was evaluated using two distinct simulation datasets and an empirical application. The first simulation incorporates non-contiguous spatial regimes, where heterogeneity is driven both by geographical proximity and data space proximity (*n*=900). The second simulation represents purely geographical proximity, where spatial heterogeneity is a continuous function of geographic distance (*n*=900); both simulation datasets are normally distributed. Together, these simulations aim to demonstrate that M-SGWR is uniquely flexible; it can accurately recovers traditional smooth surfaces as expected in GWR/MGWR, while simultaneously capturing fragmented, non-spatial patterns that these traditional models often miss.

### 3.1. Mixed effects

This simulation is constructed so that the underlying coefficient surfaces do not uniformly exhibit spatial smoothness, which departs from the assumptions typically implicit in the MGWR framework. Specifically, spatial proximity between two locations does not necessarily imply similarity in their coefficients. As illustrated in Figure 4, the intercept and the second coefficients ($\beta_0$ and $\beta_1$) follows a relatively smooth spatial surface, whereas the remaining coefficients ($\beta_2$, $\beta_3$ and $\beta_4$) display more irregular, non-smooth surface. This mixed behavior is arguably more realistic, as spatial processes do not always evolve smoothly across space. Under this data-generating mechanism, we expect M-SGWR to more accurately recover the true coefficient patterns compared to MGWR, because it explicitly accommodates a mixed-surface structure in which similarity between locations is determined not only by physical proximity but also by attribute-based proximity, thereby capturing forms of "remote closeness" that are not spatially contiguous. Equation 27 shows response variable data generation, where $(u_i, v_i)$ are the geographic coordinates for location *i*, and $\varepsilon_i \sim N(0, \ 0.9^2)$.

$$y_i = \beta_0(u_i, v_i) + \sum_{k=2}^{4} \beta_k(u_i, v_i) x_{ki} + \varepsilon_i \tag{27}$$

$$\beta_0(u_i, v_i) = 1.0 + 0.8(g_0(u_i, v_i)) \tag{28}$$

$$\beta_j(u_i, v_i) = 1.0 + s_j\left(g_j(u_i, v_i) + c_1(u_i, v_i)\right) \tag{29}$$

In the data-generating process (Equations 28 and 29), $g_j(\cdot)$ represents a latent geographic component, constructed as a smooth spatial field driven solely by physical proximity, while $c_j(\cdot)$ represents a latent contextual component, derived from a continuous regime score that captures non-geographic similarity among locations. The mixing of $g_j$ and $c_j$ in each coefficient surface controls the relative influence of geographic versus attribute-based similarity in spatially varying relationships. Subsequently the predictors were generated as mixtures of smooth geographic fields, contextual similarity, and random noise, ensuring structured but non-degenerate covariates. Under this data generating process, the M-SGWR model is expected to recover coefficient surfaces where $\beta_2, \beta_3, \beta_4$ are not purely smooth over geographic space, because their true coefficients are generated as mixtures of smooth geographic components and fragmented context-driven components. This simulation is designed to demonstrate if M-SGWR model is indeed capable of identifying different alpha values within the same model, as intercept and $\beta_1$ are purely geographic ($\alpha = 1$), while for other predictor the $\alpha$ should be within the range of zero and one.

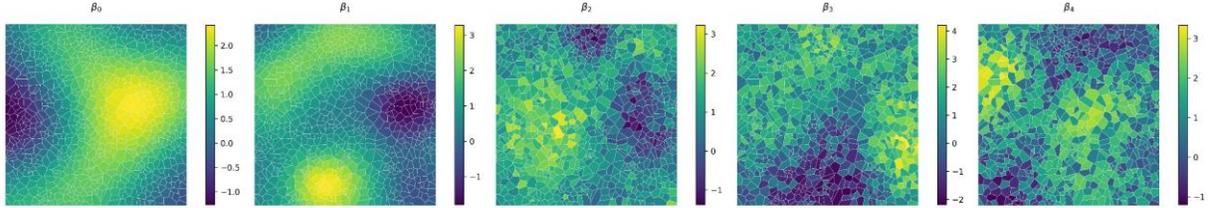

**Figure 4.** Mixed effects coefficients.

3.2. Pure geographic effect

In the second case, the coefficients are generated to follow smooth spatial surfaces driven purely by geographic proximity. The purpose of this simulation is to evaluate whether the M-SGWR model can correctly identify this structure by assigning weights solely based on geographic distance, such that the optimal parameters ($\alpha_j$) are equal to one for all predictors. The generation of estimated coefficients are shown in Equations 30-32.

$$\beta_0(u_i, v_i) = 1.0 + 0.8(g_0(u_i, v_i)) \tag{30}$$

$$\beta_1(u_i, v_i) = 1.00 + g_1(u_i, v_i) \tag{31}$$

$$\beta_2(u_i, v_i) = 1.0 + (u_i + (2 + v_i)) \tag{32}$$

In these equations, $g_0$ and $g_1$ capturing smooth random spatial variation at different degrees of spatial heterogeneity. The response value generation follows Equation 27, but the predictors and coefficients are based on this simulation concept. The predictors are generated as spatially correlated random fields with added independent Gaussian noise, while the outcome includes an additional independent Gaussian noise term to represent unexplained variability. Figure 5 visualizes the generated true coefficients based on this simulation structure.

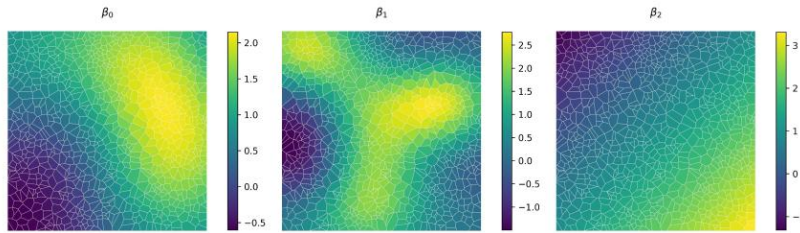

**Figure 5.** Coefficients are based on pure geographical proximity.

## 4. Results

In this section, the results of models will be compared across two simulation datasets and one empirical dataset. For the two simulation datasets, the accuracy of estimated coefficients for the simulation datasets is evaluated by the root mean squared error (RMSE) of each coefficient. Based on this evaluation metric, smaller values indicate more accurate recovery of the true local parameter values. In addition to RMSE, model performance is evaluated using multiple goodness-of-fit measures, including the adjusted coefficient of determination, the corrected AIC, and the residual sum of squares (RSS).

### 4.1. Mixed effects simulation dataset

Figure 6 compares the true coefficient surfaces (top row) with the estimated surfaces obtained from the M-SGWR (middle row) and MGWR (bottom row) models. By design, the intercept ($\beta_0$) and the first coefficient ($\beta_1$) follow smooth, geographically driven spatial patterns. Both M-SGWR and MGWR successfully recover these surfaces, producing nearly identical estimates, consistent with their shared reliance on distance-based spatial smoothing for purely geographic processes. In contrast, the remaining coefficient surfaces exhibit localized and fragmented spatial variation that departs from purely geographic smoothness. MGWR, constrained by its distance-based kernel framework, produces overly smooth estimates that fail to recover these localized deviations from the true surfaces. By comparison, M-SGWR more accurately reproduces heterogeneous spatial patterns by capturing both smooth geographic trends and localized departures, which more closely resemble spatial processes observed in real-world phenomena. These results demonstrate that M-SGWR provides greater flexibility in modeling both mixed spatial processes by jointly accounting for geographic proximity and attribute similarity and capable of identifying the pattern if it is purely based on geographical proximity, whereas MGWR tends to over-smooth coefficients when spatial variation is not exclusively governed by geographic distance.

To further assess the performance of the proposed model, Pearson's correlations between the true and estimated coefficient surfaces are presented in Figure 7. For the intercept and $\beta_1$, both models exhibit comparable correlations, reflecting their similar performance in recovering smoothly varying, geographically driven effects. In contrast, for the remaining three coefficient surfaces, M-SGWR consistently achieves higher correlations with the true coefficients, exceeding those of MGWR by 5.1, 7.0, and 12.0 percentage points, respectively.

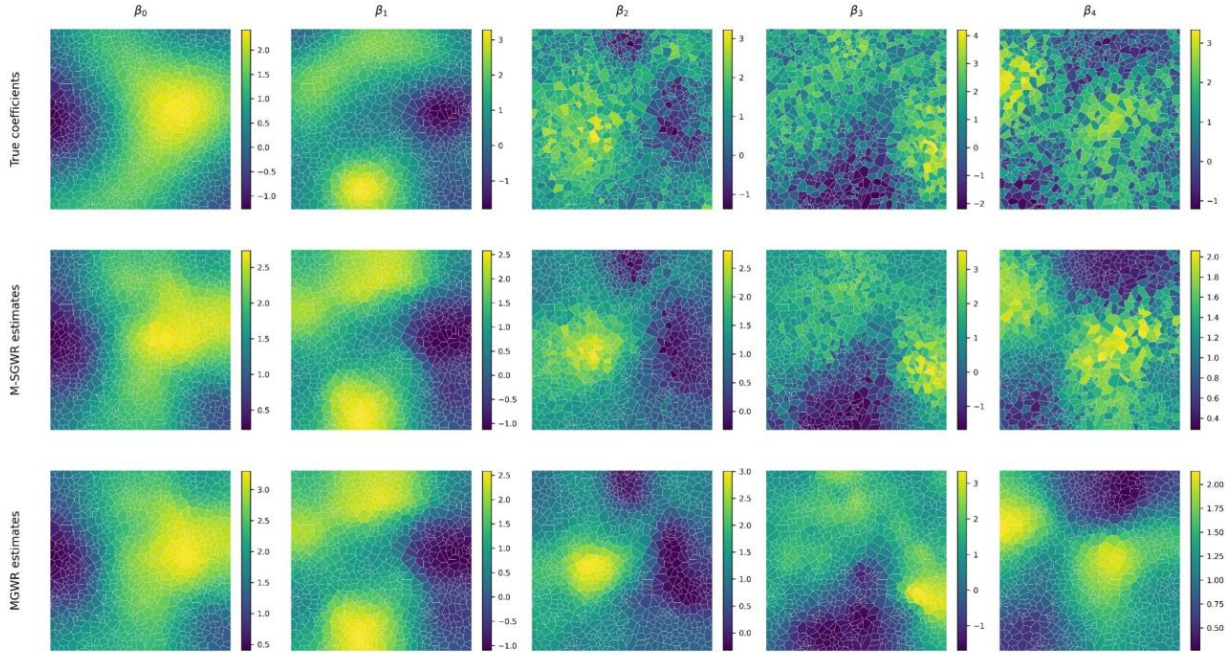

**Figure 6.** True and the estimated coefficient surfaces based on M-SGWR and MGWR models.

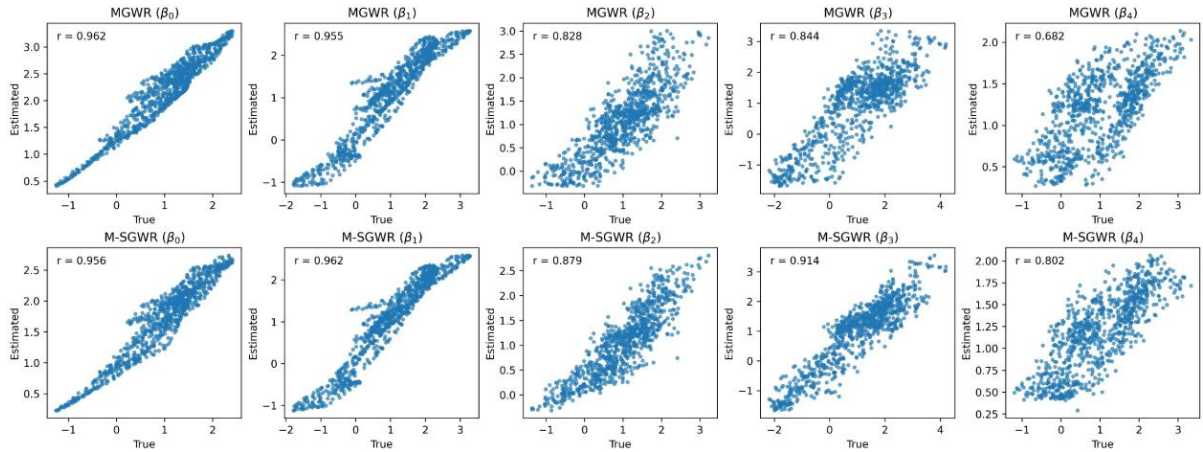

**Figure 7.** Pearson's correlation between the true and estimated coefficients of two models.

Table 1 reports the RMSE between the true and estimated coefficients. As can be observed, M-SGWR model is able to achieve lower RMSE across all coefficients, which validates that the proposed model is able to more accurately capture the true spatial patterns in the data. The M-SGWR model optimized $\alpha$ values for coefficients are 1.000, 1.000, 0.371, 0.094, and 0.240, respectively. These results indicate that $\beta_2$, $\beta_3$, $\beta_4$ exhibits the least spatial smoothness, reflecting a stronger influence of attribute similarity relative to geographic proximity. In contrast, $\beta_0$ and $\beta_1$ show a pure geographical proximity effect ($\alpha$=1.000). Alpha parameter and bandwidths will be further discussed later.

**Table 1.** RMSE between true and estimated coefficients for MGWR and M-SGWR models.

| Model  | $\beta_0$ | $\beta_1$ | $\beta_2$ | $\beta_3$ | $\beta_4$ |
|--------|-----------|-----------|-----------|-----------|-----------|
| MGWR   | 1.175     | 0.340     | 0.481     | 0.732     | 0.732     |
| M-SGWR | 0.783     | 0.313     | 0.412     | 0.561     | 0.671     |

Table 2 presents the goodness-of-fit of MGWR and M-SGWR models in addition to OLS based on mixed-effects simulation dataset. As can been across all evaluation matric, M-SGWR model outperforms MGWR model. The adjusted R-square increased from 0.864 to 0. 885 and similarly AICc and RSS of the proposed model is better than MGWR model.

**Table 2.** Models' goodness-of-fit based on mixed effects simulation dataset.

| Model  | Adj.$R^2$ | AICc      | RSS       |
|--------|-----------|-----------|-----------|
| OLS    | 0.480     | 4,025.328 | 4,553.569 |
| MGWR   | 0.864     | 2,933.779 | 928.604   |
| M-SGWR | 0.885     | 2,807.974 | 878.019   |

4.2. Purely geographical effects simulation dataset

The first simulation dataset demonstrates that the proposed model can more accurately recover the coefficient surfaces than the MGWR model by identifying spatial effects that are not solely driven by geographic proximity. Based on the second simulation dataset, in which the data generating process is governed purely by geographic proximity, M-SGWR is expected to yield results identical to those of the MGWR model. Under this setting, the optimal $\alpha$ values for all predictors should be equal to one, indicating that attribute similarity does not contribute to defining spatial relationship. After conducting the experiments, the M-SGWR model produced results identical to those of the MGWR model, as illustrated in Figure 8, where the estimated coefficient surfaces from both models are indistinguishable. These findings confirm that when attribute similarity does not contribute to the underlying process, M-SGWR appropriately converges to the MGWR solution ($\alpha = 1$), demonstrating consistency with conventional geographically weighted modeling. The Pearson correlation coefficients between the true and estimated surfaces are 0.988 for the intercept, 0.973 for $\beta_1$, and 0.988 for $\beta_2$; this result is identical to MGWR's result. The models' goodness-of-fit is reported in Table 3 for the second simulation dataset.

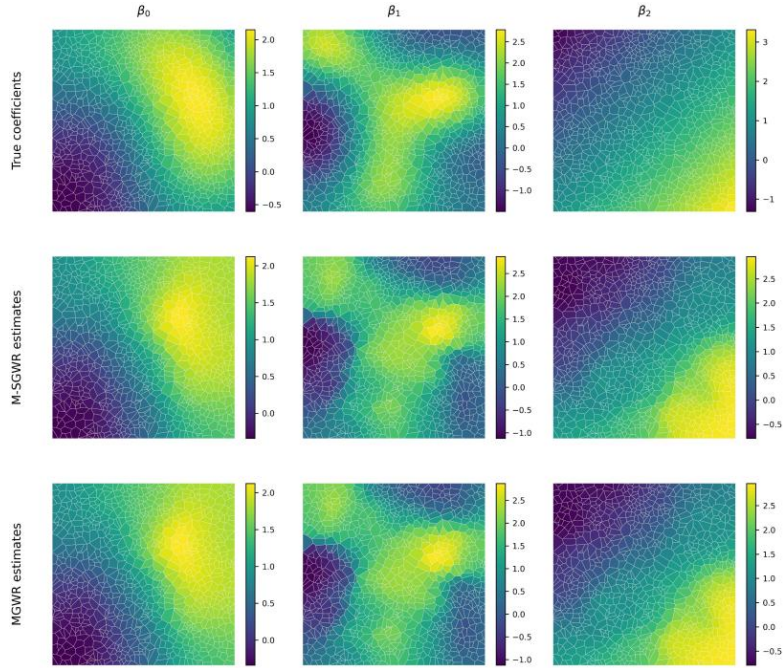

**Figure 8.** Comparison of true coefficient surfaces and estimates from M-SGWR and MGWR when spatial variation is driven exclusively by geographic proximity.

**Table 3.** Models' goodness-of-fit based on second simulation dataset.

| Model | Adj.$R^2$ | AICc | RSS |
| --- | --- | --- | --- |
| OLS | 0.435 | 3,659.600 | 3,053.590 |
| MGWR | 0.873 | 2,401.608 | 625.938 |
| M-SGWR | 0.873 | 2,401.608 | 625.938 |

### 4.3. COVID-19 data

COVID-19 case counts at the county level are used as the outcome variable for counties in Florida, Georgia, South Carolina, North Carolina, Mississippi, Tennessee, and Alabama in the United States (616 counties). The dataset includes the predictors listed in Table 4, which is sourced from this study (Stewart Fotheringham et al., 2021), however, the COVID-19 cases are sourced from Kaggle website.

**Table 4.** Predictors in COVID-19 data.

| Variable | Max | Mean | Std | Min |
| --- | --- | --- | --- | --- |
| Black population (%) | 87.20 | 24.71 | 19.95 | 0.20 |
| Hispanic population (%) | 68.50 | 6.04 | 6.59 | 0.00 |
| Holding bachelor's degree (%) | 59.80 | 19.41 | 8.97 | 3.20 |
| Median income | 112.96 | 46.32 | 11.51 | 21.50 |
| Population over 65 (%) | 56.70 | 18.41 | 4.61 | 3.20 |
| Population 18-29 (%) | 45.37 | 15.26 | 3.70 | 5.71 |
| Population density | 7.92 | 4.40 | 1.13 | 1.12 |
| Foreign born (%) | 53.70 | 4.21 | 4.65 | 0.00 |
| Uninsured (%) | 25.60 | 12.05 | 3.14 | 4.00 |

4.3.1. Model's diagnostics comparison

This section compares the performance of the M-SGWR model with OLS, GWR, SGWR, and MGWR using the evaluation metrics described in Section 4.1 based on the COVID-19 dataset. The results are consistent with the patterns observed in the simulation experiment (Table 2). The proposed model demonstrates the best performance across all metrics, followed by SGWR, while MGWR and GWR exhibit comparatively identical results.

Relative to OLS, the adjusted R-square increases substantially from 0.536 to 0.782 under M-SGWR. Compared with MGWR, the adjusted R-square improves from 0.701 to 0.782, indicating a notable gain (11.6%) in explanatory power. In addition, M-SGWR yields the lowest AICc, MAE, and RMSE values among all models, reflecting superior model fit and predictive accuracy. While MGWR improves upon GWR in terms of AICc, its overall performance remains inferior to both SGWR and M-SGWR. The comparison between SGWR and M-SGWR further suggests that allowing predictor-specific $\alpha$ values enhance the model's ability to capture complex spatial patterns. These results show that incorporating attribute-based weighting improves coefficient estimation, and that performance can be further enhanced through variable-specific optimization of the similarity parameter ($\alpha$).

**Table 5.** Models' goodness-of-fit for COVID-19 data.

| Model | Adj.$R^2$ | AICc | RSS | MAE | RMSE |
|---|---|---|---|---|---|
| OLS | 0.536 | 1,286.252 | 281.571 | 0.406 | 0.676 |
| GWR | 0.707 | 1,154.919 | 146.054 | 0.283 | 0.487 |
| SGWR | 0.747 | 1,146.136 | 115.822 | 0.251 | 0.443 |
| MGWR | 0.701 | 1,093.488 | 162.655 | 0.298 | 0.513 |
| M-SGWR | 0.782 | 943.182 | 121.397 | 0.244 | 0.433 |

Figure 9 presents the spatial distribution of locally estimated coefficients for three predictors: percentage Hispanic, population density, and median household income—derived from the MGWR and M-SGWR models. The corresponding optimized $\alpha$ values for these predictors are 0.143, 0.024, and 0.205, respectively; the joint optimization of bandwidths and $\alpha$ parameters is discussed in Section 4.2.2.

As expected under the MGWR framework, the estimated coefficients exhibit relatively smooth spatial surfaces with gradual transitions across neighboring areas, reflecting the model's reliance on geographic proximity to characterize spatial non-stationarity. In contrast, the M-SGWR estimates display greater local variability and less spatial smoothness, indicating that coefficient similarity is influenced not only by physical distance but also by other kinds of proximity (e.g., social network). Despite sharing broadly similar regional patterns, particularly for percentage

Hispanic and population density, the M-SGWR results reveal finer-scale heterogeneity that is not captured by MGWR. This suggests that geographically proximate locations do not necessarily exhibit similar local relationships with the outcome variable when data information is considered.

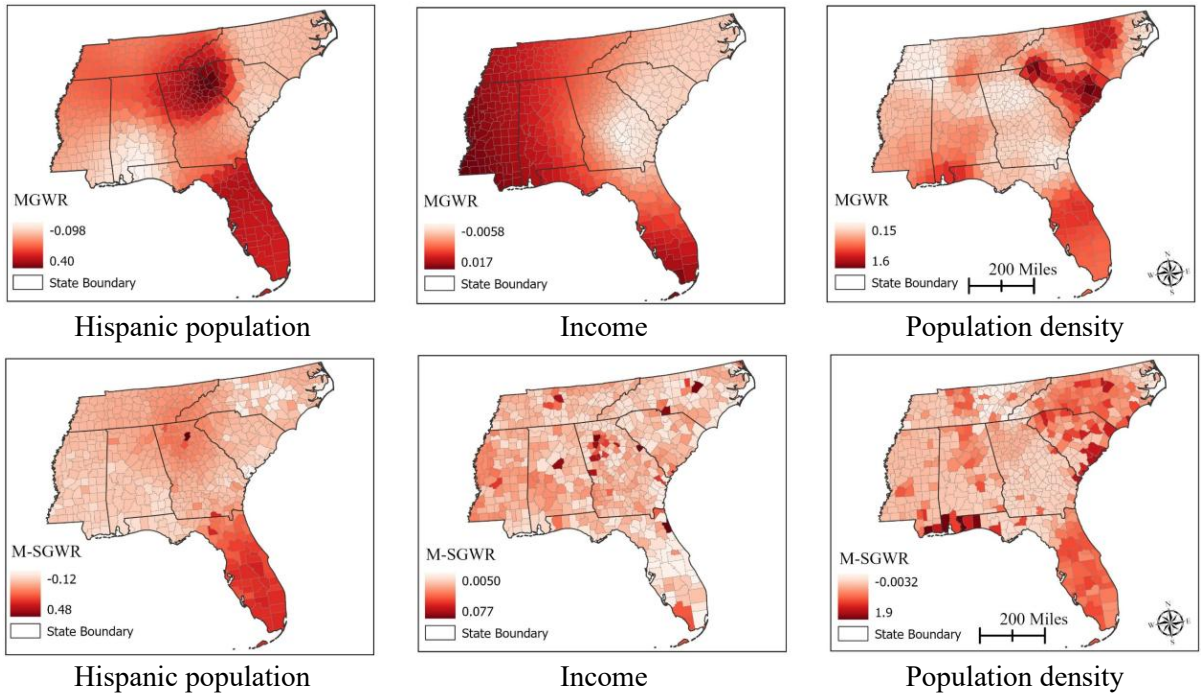

**Figure 9.** Estimated coefficients based on the two multiscale models.

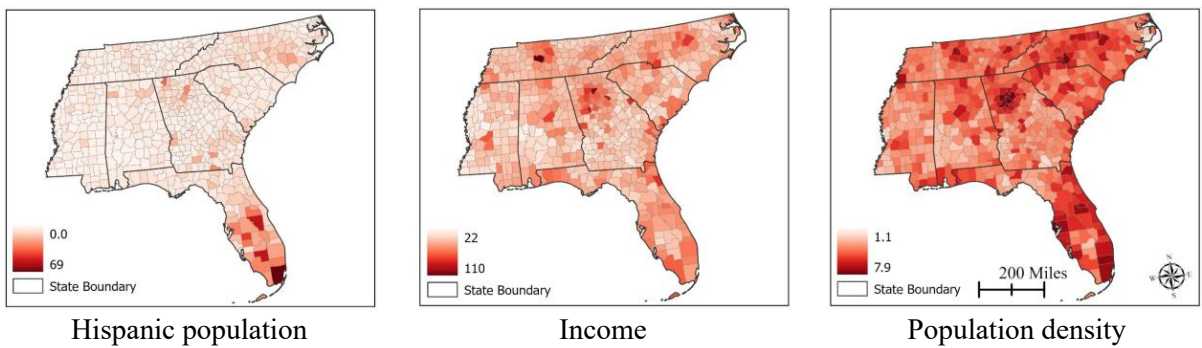

**Figure 10.** Raw values of the three predictors.

As an additional illustration, Figure 10 presents the raw values of the three predictors to provide context for the spatial variation observed in their estimated effects. Although the spatial distribution of coefficients is not expected to exactly replicate the raw predictor surfaces, their effects should reasonably reflect underlying attribute patterns. Traditional local models, such as GWR and MGWR, impose spatial smoothness by construction, encouraging neighboring locations to exhibit similar coefficients even when such similarity may not hold in practice. In contrast, M-SGWR relaxes this constraint by incorporating attribute-based information into the weighting process, allowing the model to identify spatial patterns that are not solely governed by geographic proximity.

The Global Moran's I values for the different models are reported in Table 6. Both SGWR and M-SGWR exhibit lower Moran's I values than GWR and MGWR, indicating reduced residual spatial autocorrelation. Notably, MGWR performs worse than GWR in this respect, -0.084 and -0.048, respectively.

**Table 6.** Residuals' Morna's I across models.

| Model | Morna's I | E[I] | p-value |
| --- | --- | --- | --- |
| OLS | 0.127 | -0.002 | 0.001 |
| GWR | -0.048 | -0.002 | 0.016 |
| SGWR | -0.041 | -0.002 | 0.035 |
| MGWR | -0.084 | -0.002 | 0.001 |
| M-SGWR | -0.041 | -0.002 | 0.041 |

4.2.2. Bandwidth and alpha values optimization

Table 7 presents the optimal bandwidths and alphas for the three datasets under different models using adaptive bi-square kernel function. An ($\alpha$) of close to zero indicates that the weight matrix for that variable relied heavily on attribute similarity, whereas an ($\alpha = 1$) implies a purely geographic weight structure. Intermediate values, such as ($\alpha = 0.371$) for $\beta_2$ in the first simulation, indicate that both spatial proximity (~37%) and attribute information (~63%) jointly contribute to capturing the underlying spatial process for that variable. Therefore, the M-SGWR model offers greater flexibility in optimizing the weight matrix by jointly accounting for geographic distance and attribute similarity, unlike MGWR, which relies solely on geographic proximity. This highlights that spatial interaction cannot always be fully captured by geographic distance alone; incorporating data information provides a more nuanced and accurate representation of the true spatial interactions present in the data. In the COVID-19 case study, predictors such as the percentage of Hispanic population, median income, population aged 65 and over, population density, and the proportion of foreign-born residents demonstrate that geographic proximity alone is insufficient to capture their underlying effects on COVID-19 cases. These relationships are influenced not only by spatial closeness but also by similarity in socioeconomic and demographic contexts. It is also important to note that the optimal $\alpha$ value for a given predictor is not fixed and may vary across different outcome variables. For instance, if crime rate were used as the outcome instead of COVID-19 cases, the optimal $\alpha$ associated with the Hispanic population could differ, in a manner analogous to changes in spatial scale across applications.

It is noteworthy to mention that the optimal bandwidth for the intercept can differ between the MGWR and M-SGWR models, even when the alpha in M-SGWR is set to one (section 2.2). This occurs because partial residuals influence subsequent iterations during the backfitting stage and hence leads to different bandwidth for intercept even when alpha is set to one during bandwidth optimization (Table 6). Similarly, even if an $\alpha$ of one is optimized for a predictor, this does not imply that it will share the same bandwidth in the MGWR and M-SGWR models. For example, the percentage of uninsured exhibits different optimal bandwidths across the two models (Table 6) even though the optimal alpha value is one.

**Table 7.** Bandwidths and alpha ($\alpha$) values across different models.

| Datasets | Predictors | Bandwidth (MGWR) | Bandwidth (M-SGWR) | Alpha (M-SGWR) | Bandwidth (GWR and SGWR) | Alpha (SGWR) |
|---|---|---|---|---|---|---|
| First simulation | Intercept | 105 | 105 | 1.000 | 102 | 0.994 |
| | $\beta_1$ | 90 | 90 | 1.000 | | |
| | $\beta_2$ | 77 | 78 | 0.371 | | |
| | $\beta_3$ | 44 | 56 | 0.094 | | |
| | $\beta_4$ | 130 | 124 | 0.240 | | |
| Second simulation | Intercept | 162 | 162 | 1.000 | 88 | 1.000 |
| | $\beta_1$ | 60 | 60 | 1.000 | | |
| | $\beta_2$ | 90 | 97 | 1.000 | | |
| COVID-19 | Intercept | 115 | 142 | 1.000 | 115 | 0.628 |
| | Black population (%) | 615 | 615 | 1.000 | | |
| | Hispanic population (%) | 179 | 178 | 0.143 | | |
| | Holding bachelor's degree (%) | 148 | 48 | 1.000 | | |
| | Median income | 615 | 615 | 0.205 | | |
| | Population over 65 (%) | 615 | 615 | 0.052 | | |
| | Population 18-29 (%) | 615 | 615 | 1.000 | | |
| | Population density | 43 | 43 | 0.024 | | |
| | Foreign born (%) | 490 | 544 | 0.024 | | |
| | Uninsured (%) | 481 | 615 | 1.000 | | |

Figure 11 illustrates the optimization of alpha values for coefficient surfaces in the mixed-effects simulation. The first column corresponds to $\beta_1$ and shows convergence of the optimal $\alpha$ toward one. In contrast, the remaining coefficients exhibit optimal $\alpha$ values lying between zero and one, indicating mixed spatial processes. Additionally, pronounced reductions in AICc are observed across the evaluated $\alpha$ values, highlighting the sensitivity of model fit to the balance between geographic proximity and attribute similarity.

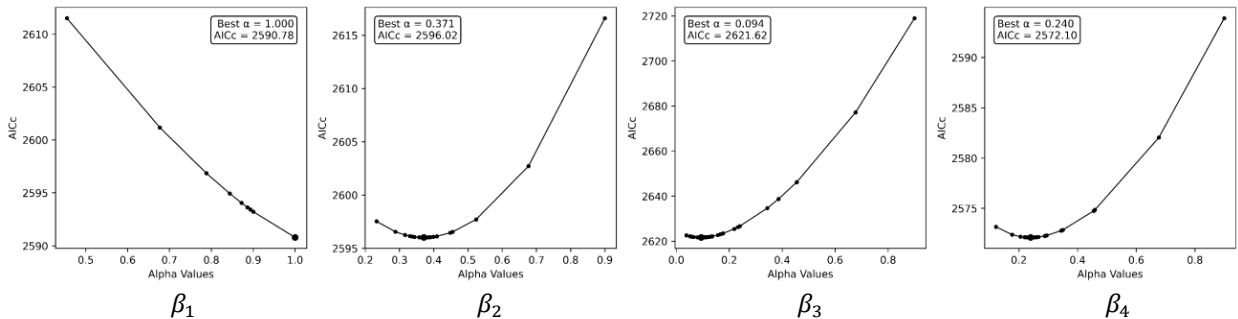

**Figure 11.** Alpha value optimization across variables for mixed-effect simulation dataset.

## 5. Discussion

### 5.1. Spatial interaction

Tobler's First Law of Geography (Tobler, 1970) is one of the foundations of spatial analysis, which states that everything is related, but nearby things are more strongly related than distant ones. In this definition, the term 'near' and 'related' are the two core components of the law, and these two terms can be interpreted as how similar or dissimilar two locations are (Anselin & Li, 2020). Yet, their definitions are ambiguous in the context of spatial analysis on how to quantify 'near' and 'related.' Anselin and Li translated 'near' to geographic similarity and 'related' to attribute similarity. Local regression models however, such as GWR and its extensions implement the law by translating proximity solely into kernel weights derived from a physical distance metric (e.g., Euclidean or road network) (He et al., 2023; Lu et al., 2019; Oshan et al., 2019). This distance-based weighting is a convenient proxy for spatial relationships. In practice, the notions of 'near' and 'related' are much more complex, which cannot be described purely by geographic proximity.

We argue that relying solely on geographic distance presents two key limitations. (1) Spatial interaction is a complex process and cannot be determined only by physical proximity. In a globalized and digitally connected world, functional connectivity that can be captured by flows of people, goods, information, or capital, can create 'remote closeness'. Flow- or mobility-based weight matrices can assign high relatedness to farther regions with strong exchange (e.g., intensive air travel or trade), and low relatedness to adjacent regions with weak interaction (e.g., limited cross-border movement or divergent institutional ties). (2) Spatial interaction is a dynamic process, whereas physical distance is static. Relationships between places can strengthen or weaken over time due to numerous reasons such as policy changes, economic cycles, or social factors; consequently, weights that measure spatial interaction need to be dynamic to reflect evolving dependence. For example, when modeling COVID-19 cases in 2020 using M-SGWR model, the optimal alpha for Hispanic population is 0.143 with respect to other predictors, while in another year for the same location for this predictor, the optimal alpha can differ. This flexibility allows the model to capture dynamic spatial dependencies over time.

Incorporating a new dimension alongside geographical space to quantify spatial interaction in local regression analysis can provide a more logical and comprehensive framework for identifying complex spatial patterns. Lessani and Li proposed extending the measurement of spatial relationship by introducing data space (attribute space) as a complementary dimension to geographic space (SGWR) (Lessani & Li, 2024). The SGWR model can indicate whether spatial relationships exhibit mixed effects or are purely geographically driven; however, it cannot identify which specific predictors exhibit such behavior (Table 7). This research demonstrates that extending SGWR to a multiscale framework (M-SGWR) further enhances the model's explanatory power. In M-SGWR, each predictor is assigned a distinct mixing parameter ($\alpha$) and geographic scale, enabling optimized, variable-specific characterization of spatial interactions between the same pair of locations.

We also acknowledge, for phenomena governed primarily by physical processes with distance-decay, geographic distance remains an appropriate primary determinant of spatial interaction, as those variables would adopt and $\alpha = 1$ (second simulation dataset), which refers that physical distance is the primary measurement of spatial interaction in the model. Thus, a comprehensive approach therefore treats 'near' and 'related' as measurable constructions that may combine geographic proximity with functional connectivity, rather than equating relatedness with geographical distance alone. Here, functional connectivity refers to the degree of interaction, exchange, or influence between two locations, regardless of their physical proximity.

5.2. M-SGWR as a unified model framework

The M-SGWR model can be regarded as a unified framework that encompasses the GWR, SGWR, and MGWR models. Depending on the specification of bandwidths and alpha parameters, the M-SGWR model can operate in equivalence to other models, as follows:

- GWR equivalence: In GWR, a single bandwidth is applied, meaning that all processes are assumed to operate at the same spatial scale. M-SGWR reduces to GWR when the optimal bandwidths for all variables are equal ($bw_{x0} = bw_{x1} = ... = bw_{xn}$) and the alpha values for all variables are one ($\alpha_{x0} = \alpha_{x1} = ... = \alpha_{xn} = 1$).
- SGWR equivalence: SGWR maintains the concept of a single spatial scale, as in GWR, but incorporates an attribute weight matrix controlled by a global alpha parameter. Thus, M-SGWR behaves as SGWR when the optimal bandwidths are identical across all variables ($bw_{x0} = bw_{x1} = ... = bw_{xn}$) and the alpha values are the same for all variables too but ranges between zero and one ($\alpha_{x0} = \alpha_{x1} = ... = \alpha_{xn}, \alpha \in [0,1]$).
- MGWR equivalence: MGWR allows each variable to operate on its own spatial scale ($bw_{x0} \neq bw_{x1} \neq ... \neq bw_{xn}$). M-SGWR is equivalent to MGWR when the optimal bandwidths match those of MGWR, and the alpha values are all equal to one ($\alpha_{x0} = \alpha_{x1} = ... = \alpha_{xn} = 1$), as also proven by the second simulation dataset.
- Full M-SGWR: When neither of the above conditions are satisfied that is when M-SGWR optimizes a distinct bandwidth for each variable and assigns distinct alpha values and the model functions as a true multiscale approach in both geographical space and attribute space; shown in the first simulation and housing datasets.

Future studies could focus on developing adaptive mechanisms that automatically identify appropriate model structures (single-scale vs. multiscale) and plausible $\alpha$ ranges based on data characteristics without model calibration, thereby reducing computational cost and improving model applicability.

5.3. Computational time

Local regression models often suffer from low computational efficiency because a separate local model is generated for each observation. The M-SGWR model, like other local regression frameworks, is not exempt from computational inefficiency. For MGWR, the backfitting stage has a time complexity of $O(kdn^2 \log n)$ and the inference stage $O(kdn^3)$ when using Golden Section Search, where $k$ is the number of variables, $d$ is the number of backfitting iterations until convergence, and $n$ denotes the number of observations. In M-SGWR, there are two additional steps: $\alpha$ optimization and attribute weight matrix computation, which increase the backfitting cost compared to MGWR. Specifically, the backfitting complexity becomes $O(kdn^2 \log n \log(\frac{1}{\varepsilon}))$ when using a divide-and-conquer alpha search with precision $\varepsilon$, and $O(kdn^2 \log n \, (\frac{1}{h}))$ when using a greedy hill-climb search with fixed step size $h$. In the inference stage, the $\alpha$ value is already optimized (like the bandwidth), so only an attribute weight matrix needs to be computed for each observation, analogous to the spatial weight matrix. This adds $O(kn^2)$ to the cost, which is negligible compared to the $O(kn^3)$ complexity of matrix multiplications. Therefore, the inference-stage complexity of M-SGWR is the same as MGWR, though slightly slower in practice due to the extra attribute weight calculation. To accelerate M-SGWR in terms of computation time, the concept of parallel computing can be applied to both the backfitting and inference stages similar to other models (e.g., FastGWR/MGWR and FastSGWR). Additionally leveraging parallel processing will improve the model's efficiency, enhance its usability, and make it feasible for application with larger spatial datasets.

## 6. Conclusion

This study introduces a new local regression framework, M-SGWR, which integrates spatial and attribute weight matrices within a multidimensional context. In M-SGWR, each predictor is assigned a distinct pair of weight matrices, one capturing geographic proximity and the other attribute similarity, which are then combined through an optimized parameter ($\alpha$) for model estimation. This new framework enhances our ability to more accurately measure spatial dependency in local spatial regression models. Unlike GWR or MGWR, which constrain the measurement of spatial interaction solely to geographical distance, the proposed model captures how each spatial process is differentially influenced by geographic proximity and what the data itself reveals about the spatial relationship underlying the process. This allows the model to capture complex spatial patterns with greater accuracy than MGWR and GWR models.

The proposed model was evaluated using two simulation datasets and one empirical dataset; the results demonstrate that spatial interaction cannot be adequately captured solely by geographical distance. Compared to MGWR, an advanced multiscale extension of GWR, the M-SGWR model achieved higher goodness-of-fit across all datasets. These results suggest that contemporary societal interactions are far more complex, and geographical distance alone no longer represents how spatial processes function. They further indicate that "remote closeness" has become an integral component of modern spatial dynamics, influencing some processes more strongly than others. The M-SGWR model thus captures this complexity by leveraging information inherent in the data itself to measure how places interact, in addition to their geographical proximity.

Despite the substantial improvements over the MGWR model, we acknowledge several limitations of M-SGWR that present opportunities for future research. First, alternative strategies for bandwidth and alpha optimization could be developed to replace the iterative backfitting procedure, such as simultaneous optimization of both parameters. Additionally, the backfitting algorithm may encounter difficulties with local minima; therefore, alternative metaheuristic methods such as simulated annealing or genetic algorithms could be employed to explore the global search space, although this would come at the cost of increased computational time. Second, because regression models are sensitive to data characteristics, M-SGWR should be applied to a broader and more diverse range of datasets to further evaluate and validate its performance; also, different types of kernel functions should be examined in addition to adaptive bi-square kernel. Finally, currently, M-SGWR optimizes bandwidth and $\alpha$ using AICc and CV. Future work could incorporate alternative criteria (e.g., RSS).

**Data and code availability statement**

The datasets and codes used for this study are publicly available here:
https://github.com/Lessani252/M-SGWR